\documentstyle[a4,12pt,epsfig,psfig]{article}
\renewcommand{\baselinestretch}{1.0} 
\begin{document}

\renewcommand{\baselinestretch}{0.7} 
\begin{flushright}
\Large {VECC - NEX / 97003}
\end{flushright}
\vskip 1.0cm
\begin{center} 
{\Large {\bf {Experimental search for DCCs using Robust Observables}}}
\renewcommand{\baselinestretch}{1.0} 

\vskip 0.6cm

Debsankar Mukhopadhyay{\footnote{E-Mail:
dsm@vecdec.veccal.ernet.in}}           {\footnote{Financially supported
by Council of Scientific and Industrial Research, Ministry of Human
Resource Development Group, Government of India.}}   {\footnote{Talk
presented at the 3rd International Conference on Physics and
Astrophysics of Quark-Gluon Plasma, March 17 - 21, Jaipur, India.}}

\small {\em for WA98 Collaboration} \\
\small {\em Variable Energy Cyclotron Centre,}\\
\small {\em 1/AF, Bidhan Nagar, Calcutta - 700064, }\\
\small {\em India.}
\end{center}
\renewcommand{\baselinestretch}{1.0} 
\vskip -0.5cm
\begin{abstract}
\small{
\small{
We have analyzed the joint charged particle to photon density
correlations in WA98 experiment at CERN-SPS in order to study the
formation of the chirally disoriented vacuum in Pb + Pb collisions at
158 AGeV. 
The method of analysis is based on the event-by-event estimation of  a 
set of robust observables which are the ratios of normalized
bivariate factorial moments corrected for many efficiency factors.
This is found to be 
a very powerful technique in identifying, within the limits of the 
detector resolution, the events with the localized charged 
particle to photon density fluctuations. }}
\end{abstract}
\vskip 1.0cm
\renewcommand{\baselinestretch}{1.0} 

In high energy heavy ion collisions, hot fireballs may be produced with
interiors having anomalous chiral order parameters. Eventually, these 
conjectured domains of disoriented chiral condensate(DCC) would decay 
by radiating pions with distinctive isospin distribution,
following an inverse square-root law {\cite
{a}} 

\renewcommand{\baselinestretch}{0.9} 
\begin{equation} 
\biggl(\frac{dP}{df}\biggr)_{DCC}= \frac{1}{2\sqrt{f}},
\end{equation}
where  $f$ is the neutral pion fraction, given by
\renewcommand{\baselinestretch}{0.7} 
\begin{equation} 
f = \frac{N_{\pi^0}} {N_{\pi^0}+ N_{\pi^+}   +   N_{\pi^-}}  \sim \frac{N_{\gamma}/2}{N_{\gamma}/2 + N_{ch}}
\end{equation}
\renewcommand{\baselinestretch}{0.8} 
This is markedly different from the generic distribution, 
which is governed by a binomial distribution sharply peaked at $f
\approx 1/3$.
\renewcommand{\baselinestretch}{1.0} 

Therefore, an experimental  study of event by event charged to neutral density
correlation would shed light on this exotic phenomenon.

WA98{\cite {wa98}}
is the only heavy ion experiment at CERN SPS which has initiated 
the `vacuum engineering' in Pb + Pb collisions at 158 AGeV. 
The multiplicities of the charged particles are measured by a Silicon
Pad Multiplicity Detector(SPMD) within a pseudorapidity coverage of
$2.35 \leq \eta \leq 3.75$. Photon multiplicities are measured in the
range of $2.4 \leq \eta \leq 4.4$ by a highly granular Photon
Multiplicity Detector(PMD).  The efficiency of detecting a charged
particle in the active area of the detector is $\sim 99\%$, while 
the average photon counting efficiency is $70\% - 80\%$ depending on
centrality, with a $35\%$ contamination of showering hadrons.
The centrality of the collision is determined by a Mid Rapidity
Calorimeter (MIRAC) and a Zero Degree Calorimeter(ZDC) by respective
measurement of the transverse and forward energies.

In addition to the
detection efficiencies for charged particles and photon measurements, the
following probability factors also exist because of finite acceptance of
the detectors: 

(a) $\epsilon_{ch}$: Probability of observing a $\pi^{\pm}$ within the detector
acceptance,
(b)  
$\epsilon_0$: Probability that none of the photons from a $\pi^0$ falls
in the detector,
(c) $\epsilon_1$: Probability that one of the $\gamma$'s from a $\pi^0$
comes within the detector coverage,
(d) $\epsilon_2$: Probability of observing both the $\gamma$'s from a
$\pi^0$ within the detector acceptance.

These $\epsilon$ factors along with the detection efficiencies and the
background effects{\cite {nayak}} put some serious challenges for us in
this game.

Nevertheless, a  sophisticated analysis technique, based on the 
method of the normalized
factorial moments, has been developed {\cite {min}} to estimate, within
limits, the inclusive charged to neutral density correlations.
In this formalism, the acceptance factors, $\epsilon_{ch}$ 
and $\epsilon_i$ (i=0,1,2), which can not be directly estimated from the
data, are incorporated in the  generating function of the bi-variate factorial 
moments. The resulting factorial moments contain the
efficiency factors. 

It can be shown{\cite {min}} that the proper combinations of these 
factorial moments eliminate the
acceptance factors and formulate the following `robust' observables {\cite
{min}}:
\renewcommand{\baselinestretch}{0.8} 
{\small{
\begin{equation}
r_{i,1} = \frac{F_{i,1}}{F_{i+1,0}},
\end{equation}
where $F_i$ and $F_{i,j}$ are given by: 
\begin{equation}
F_{i} \equiv \frac{\langle N(N-1)\ldots (N-i+1)\rangle} {\langle
N\rangle ^{i}}  
\end{equation}
\begin{equation}
F_{i,j}=\frac{\left< n_{ch}(n_{ch}-1)\ldots
(n_{ch}-i+1)~n_{\gamma}(n_{\gamma}-1) \ldots (n_{\gamma}-j+1)\right>}
{\left<n_{ch}\right>^{i}\left<n_{\gamma}\right>^{j}}\ .
\end{equation}
}}
\renewcommand{\baselinestretch}{1.0} 
In the equations cited above, the angular brackets $<...>$ denote the 
averages over a large number of
events and $N$ is the individual event multiplicity.

Moreover, one finds that for all $i \geq 1$
{\small {
\renewcommand{\baselinestretch}{0.7} 
\begin{eqnarray}
r_{i,1}(generic) &=& 1, \nonumber \\
r_{i,1}(DCC) &=& \frac{1}{i+1}  
\end{eqnarray}
in the two cases.
}}
\renewcommand{\baselinestretch}{1.0} 

The above formalism developed by the MiniMax Collaboration is not
sensitive to the localized fluctuations in single events. In order to
make it suitable to identify the single event fluctuation structures, we
have to calculate the robust observables on an event-by-event basis. In
that case the angular brackets in the numerators and denominators 
of equations (4) - (5) should denote the averages over a considerable
number of bins in a single event phase space distributions of $N_\gamma
- N_{ch}$. Now, the question comes whether one can use equation (6) to
distinguish between a pure DCC and pure Generic event. The issue here is
the size and correlation length of a DCC domain {\cite {tay}}. For
example,
if there is a single domain covering the  entire detector acceptance, then one
would
expect that each bin (up to statistical fluctuations) would have the
same
orientation.
On the other hand, if the DCC were produced throughout the acceptance
but
the correlation length is comparable to the bin size, then one might
expect a
$1/\sqrt{f}$ distribution (contaminated by whatever background is present)
more
or less bin by bin and in that case equation (6) can be used to
distinguish a DCC and a generic event. The purpose of the
multi-resolution analysis with moments or wavelets is to find out this
type of correlations. Hence, in the WA98 environment we shall compute
the robust observables event - by - event and try to explore their
properties. This is important for using them as a distinctive tool for a
DCC and generic event.

To analyze the data, we have selected the central events  
which correspond to about top 10$\%$ of the minimum bias 
cross-section. The selection of the central events confine us to the 
high  detection efficiency regions specially for photons. The 
joint charged particle / photon distributions were constructed by using 
the  overlap region $3.25 \leq \eta \leq 3.75$ of PMD and SPMD 
for the full azimuthal coverage.
We have partitioned the whole azimuthal coverage  
into four different types of  bins, namely,
(a) 32 bins: $\Delta \phi$ = $11.25^\circ$, 
(b) 16 bins: $\Delta \phi$ = $22.50^\circ$, 
(c)  8 bins: $\Delta \phi$ = $45.00^\circ$, 
(d)  4 bins: $\Delta \phi$ = $90.00^\circ$,
in order to achieve the multi-resolution binning of the available
$\eta - \phi$ region.

In order to search for any anomalous charged to neutral particle density
fluctuation, we have computed the robust observables for the following
three sets of events:
(1) generic events,
(2)  simulated exotic
events and (3) experimental data.

The generic events are generated by the Monte-Carlo code VENUS
{\cite{wer}}. The detector effects are introduced in VENUS in order to
generate data-like generic events. 

The simulation of the exotic events
has been done by modifying the output of the VENUS event generator to
include the fluctuations in the relative production of the charged
particles and neutral pions. We assume that a single DCC domain
translates to the percent of VENUS pions that gets affected. The
identities of these affected pions are altered by choosing a certain
value of $f$ randomly from the distribution given by (1). For the
present purpose,
we have simulated the exotic events in such a way that all of the pions
within $3.0\leq\eta\leq4.0$ and $0^{\circ}\leq\phi\leq45^{\circ}$
are supposed to come from a DCC domain. This is an example of a very
large domain of DCC which may not be created in experiments but is very
useful to test the sensitivity of the method towards the  localized charged
to neutral fluctuations.

\renewcommand{\baselinestretch}{1.0} 

The results of the analysis are summerized in Table -1 for
the above three cases. The entire analysis has been done with a large
number of events so that the statistical errors are small.
 In case of the data and simulated generic events,
the distributions of $r_{i,1}$ are peaked around 1.00. But the mean value 
of the distributions shift to a lower value in case of the simulated 
exotic events. This indicates the sensitivity of the method to
disentangle exotic events accompanied with a large domain of DCC from
the generic samples. Moreover, the difference of the data and the simulated
exotic events indicates the probability of the formation of smaller
and/or multiple domains of DCC. Further simulations in this direction are
in progress. The sensitiveness of the higher order moments (say,
$r_{3,1}$) to the localized fluctuations is remarkable within the limits
of the so called `empty bin effects'. The root mean square values of the
distribution of $r_{i,1}$, for all resolutions, are higher for data than
the simulated normal events. Also, the values of $\sigma$ increases with
decreasing bin size. This indicates that one should not go up to the
smallest bin samples to avoid the trivial statistical effects.

\renewcommand{\baselinestretch}{0.8} 
\vskip 0.6cm
\begin{center}
Table - 1
\end{center}

The average values of $r_{i,1}$ for the simulated normal  events, simulated
exotic events and
data. 
\vskip 0.2cm
\begin{center}
\small {

\begin{tabular}{|l|l|l|l|l|}\hline
Event type&No. of Bins&$r_{1,1} \pm \sigma$&$r_{2,1} \pm \sigma$&$r_{3,1} \pm
\sigma$\\\hline
 Normal&4&$1.001 \pm 0.02$&$0.997 \pm 0.02$&$1.008 \pm
0.03$\\
Exotic&&$0.845 \pm 0.04$&$0.806 \pm 0.04$&$0.742 \pm 0.05$\\
Data&&$1.009 \pm 0.02$&$1.019 \pm 0.03$&$1.031 \pm 0.05$\\\hline\hline
 Normal&8&$1.001 \pm 0.01$&$1.004 \pm 0.02$&$1.009 \pm
0.04$\\
Exotic&&$0.852 \pm 0.04$&$0.848 \pm 0.04$&$0.801 \pm 0.05$\\
Data&&$1.012 \pm 0.03$&$1.027 \pm 0.05$&$1.044 \pm 0.07$\\\hline\hline
 Normal&16&$1.002 \pm 0.02$&$1.007 \pm 0.04$&$1.010 \pm 0.06$\\
Exotic&&$0.889 \pm 0.04$&$0.853 \pm 0.04$&$0.831 \pm 0.05$\\
Data&&$1.021 \pm 0.05$&$1.046 \pm 0.07$&$1.080 \pm
0.11$\\\hline\hline
 Normal&32&$1.002 \pm 0.04$&$1.009 \pm 0.08$&$1.090 \pm
0.12$\\
Exotic&&$0.901 \pm 0.04$&$0.889 \pm 0.05$&$0.821 \pm 0.08$\\
Data&&$1.028 \pm 0.05$&$1.067 \pm 0.11$&$1.130 \pm 0.19$\\\hline
\end{tabular}
}
\end{center}

A closer look into the data indicates a few events
having the values of $r_{i,1} > 3\sigma$, where $\sigma$ denotes the
root mean square value of the concerned distribution.
These events
may be interesting with the indication of considerable charged
particle / photon density fluctuations in the azimuthal space. But a
concrete inference needs more studies on the detector effects as the
data
of both the PMD and SPMD are in a preliminary stage.
This is, therefore, evident that in the
first attempt of event-by-event computation of  the robust observables,
we have established the  suitability of the moment analysis  in the
DCC search.  It is found that the data contains more
fluctuations than the simulation and for both the data and simulation
$\sigma$ increases with decreasing bin size.
Also, the preliminary data indicates the presence of remarkable 
charged to neutral
density fluctuations in some events. These results, thus, demand a more
careful analysis of the detector effects for a confident conclusion.

\renewcommand{\baselinestretch}{0.8} 

\end{document}